\documentclass[aps,twocolumn,preprintnumbers,superscriptaddress,floatfix,nofootinbib,showpacs]{revtex4-1}
\usepackage{graphicx} 
\usepackage{amsmath,amssymb}
\usepackage{subfigure}
\usepackage{multirow} 
\usepackage[colorlinks,linkcolor=blue]{hyperref}
\usepackage{color}
\usepackage{newtxtext, newtxmath}
\usepackage{braket,bm}
\usepackage{enumitem}
\usepackage{epstopdf}
\newcommand*{\slashed}[1]{{#1\!\!\!/}}
\newcommand*{\hc}{\text{H.\,c.}}
\pdfpagewidth=\paperwidth
\pdfpageheight=\paperheight
\allowdisplaybreaks

\begin{document}

\title{Analysis of the data on differential cross sections and spin density matrix elements for $\gamma p \to \rho^0 p$}

\author{Ai-Chao Wang}
\affiliation{School of Nuclear Science and Technology, University of Chinese Academy of Sciences, Beijing 101408, China}
\affiliation{College of Science, China University of Petroleum (East China), Qingdao 266580, China}

\author{Neng-Chang Wei}
\affiliation{School of Physics, Henan Normal University, Henan 453007, China}

\author{Fei Huang}
\email[Corresponding author. Email: ]{huangfei@ucas.ac.cn}
\affiliation{School of Nuclear Science and Technology, University of Chinese Academy of Sciences, Beijing 101408, China}

\date{\today}

\begin{abstract}

The newly published data on spin density matrix elements from the GlueX Collaboration, along with the previously released differential cross-section data from the CLAS Collaboration and the other two experiments for the $\gamma p \to \rho^0 p$ reaction, are systematically investigated using an effective Lagrangian approach within the tree-level Born approximation. The model combines contributions from $t$-channel meson exchanges ($\pi$, $\eta$, and $f_2$), $s$-channel nucleon ($N$) and nucleon resonance ($N^\ast$) exchanges, $u$-channel $N$ exchange, and a generalized contact term to construct the scattering amplitudes. Regge propagators are employed for $t$-channel amplitudes to incorporate the contributions from mesons with various spins lying on the same trajectories. The analysis shows that the background contributions, dominated by the $f_2$-trajectory exchange, provide a satisfactory description of the data in the high-energy and forward-angle regions. The inclusion of specific nucleon resonances, such as $N(2100)1/2^{+}$, $N(2060)5/2^{-}$, or $\Delta(2000)5/2^{+}$, significantly improves the description of the differential cross-section data at near-perpendicular scattering angles in the low-energy region. Predictions of photon beam and target nucleon asymmetries are provided, offering valuable insights to discriminate reaction mechanisms when corresponding data become available in the future.

\end{abstract}



\maketitle

\section{Introduction}   \label{Sec:intro}

A comprehensive understanding of nucleon resonances ($N^\ast$'s)  is crucial for gaining insights into the nonperturbative regime of quantum chromodynamics (QCD). Traditionally, our knowledge about $N^\ast$'s has been primarily acquired through $\pi N$ scattering or $\pi$ photoproduction reactions. However, certain theoretical predictions for $N^\ast$'s, proposed by quark models \cite{Isgur:1978,Capstick:1986,Loring:2001} or lattice QCD \cite{Edwards:2011,Edwards:2013}, have yet to be verified in experiments, leading to their identification as ``missing resonances'' \cite{Mart:1999ed}. The absence of these resonances in experiments may be attributed to their weak coupling to the $\pi N$ channel, making them challenging to detect. This scenario necessitates an exploration of alternative, strongly-coupled reaction channels. Over the past few decades, meson photoproductions, including $\gamma N\to \eta N$, $\eta'N$, $\rho N$, $\omega N$, $\phi N$, $K\Lambda$, $K\Sigma$, $K^*\Lambda$, $K^*\Sigma$, $K\Lambda^*$, and $K\Sigma^*$ reaction processes, have drawn growing interest and led to important progress in both experimental and theoretical domains \cite{Kashevarov:2017,Anisovich:2018,Zhangy:2021,Bradford:2006,Mart:2019,CLAS-beam,Zachariou:2020kkb,Kim:2014,hejun:2014,Yu:2017c,Wei:2022,Wang:2017,Wang:2018,Wang:2022, Anisovich:2017rpe,Wei:2020,Moriya:2013,Wangx:2020,wangen17,kim17,Zhangyx:2021,Wein:2021,Roy:2018,Weinc:2019,Wangxiaoyun:2017}. These processes offer alternative platforms for investigating $N^\ast$'s, particularly those in the less-explored higher energy region. In the present study, we focus on the $\gamma p \to \rho^0 p$ reaction.

Experimentally, several collaborations have conducted measurements on differential cross sections and spin density matrix elements for the $\gamma p \to \rho^0 p$ reaction. Particularly noteworthy is the release of new spin density matrix element data by the GlueX Collaboration in 2023, conducted at a photon laboratory energy of $E_{\gamma}= 8.5$ GeV \cite{newsdme:2020}. Very close to this energy region ($E_{\gamma}= 9.3$ GeV), data previously published by SLAC on spin density matrix elements are also available \cite{oldsdme:1973}. Additionally, for low-energy differential cross sections, the CLAS Collaboration provided data in 2001 at three energy points ($E_{\gamma}=3.28,3.55,3.82$ GeV) \cite{Battaglieri:2001}. In 2005, Wu et al. published differential cross-section data near the $\rho N$ threshold energy region \cite{Wu:2005}. Older but valuable differential cross-section data published in 1972 \cite{Ballam:1972} are also considered in our analysis.

Theoretically, several works have been devoted into the study of $\gamma p \to \rho^0 p $ reaction. In Ref.~\cite{Laget:1995}, Laget and Mendez-Galain studied the photo- and electro-photoproduction of light vector mesons at large momentum transfer. In Ref.~\cite{Laget:2000}, Laget reanalyzed the photoproduction of $\rho^0$, $\omega$, and $\phi$ mesons, and their findings indicated that, at forward angles in the high-energy region, the cross-sections can be well accounted for by Pomeron exchange, while the contributions of $t$-channel exchange of Reggeons play a significant role at low energies. In the work by Oh et al. in Ref.~\cite{Oh:2004}, a comparison was made between the $\sigma$-exchange model with $\sigma$ coupling parameters fitted to experimental data and a model incorporating constructed 2$\pi$, $\sigma$, and $f_{2}$ exchanges for $\rho$ meson photoproduction. It was found that the latter provided a better description of the low-energy differential cross-section data. The study conducted by Riek et al. in Ref.~\cite{Riek:2009} focused on medium effects, employing a previously constructed $\rho$ propagator based on hadronic many-body theory. Additionally, within a model based on Regge theory in Ref.~\cite{Mathieu:2018}, Mathieu et al. provided a description of the SLAC data \cite{oldsdme:1973} for $\rho^0$, $\omega$, and $\phi$ photoproductions, and made predictions for the energy and momentum-transfer dependence of spin-density matrix elements at $E_{\gamma}=8.5$ GeV.

One sees that the majority of previously published theoretical works have focused solely on analyzing the available low-energy differential cross-section data for $\gamma p\to \rho^0 p$ \cite{Battaglieri:2001,Wu:2005,Ballam:1972}. But, notably, the data on spin density matrix elements newly published by the GlueX Collaboration in 2023 \cite{newsdme:2020}, which are expected to provide more constrains on the reaction amplitudes in theoretical models, have never been analyzed. It is obvious that an independent and comprehensive investigation that incorporates both the newly released spin density matrix elements data \cite{newsdme:2020} and the low-energy differential cross-section data \cite{Battaglieri:2001,Wu:2005,Ballam:1972} for $\gamma p\to \rho^0 p$ is necessary and indispensable, which will most likely offer us with a more accurate and reliable understanding of the reaction mechanism  and resonance contents and parameters for $\rho^0$ photoproduction reactions.

In the present study, we employ the effective Lagrangian method at the tree-level approximation to conduct an comprehensive investigation into the recently released spin density matrix element data \cite{newsdme:2020} and previously available low-energy differential cross-section data \cite{Battaglieri:2001,Wu:2005,Ballam:1972} for $\gamma p\to \rho^0 p$. We consider the $t$-channel $\pi$, $\eta$, and $f_2$ exchanges, $s$- and $u$-channel nucleon ($N$) exchange, a generalized contact term, and possible contributions from the $s$-channel nucleon resonances in constructing the reaction amplitudes to reproduce the data. The considered differential cross-section data are in the energy range $E_\gamma=1.5 -5$ GeV, and the spin density matrix element data are at the photon energy $E_{\gamma}=8.5$ GeV. To account for the high-energy region data, we model the $t$-channel exchanges in the Regge style.

It should be emphasized that a thorough investigation and extraction of nucleon resonances require the implementation of a coupled-channels approach \cite{Shen:2024nck,Wang:2023snv,Ronchen:2012eg,Wang:2022oof,Cao:2013psa,Fernandez-Ramirez:2015tfa,Kamano:2015hxa,Anisovich:2011fc}, which has been primarily utilized in studies of pseudo-scalar meson production. This framework preserves the unitarity and analyticity of the reaction amplitude and allows for a systematic search for resonance poles in the complex energy plane. While such an approach is indispensable for a more detailed understanding of the reaction mechanism, the present study focuses on laying the groundwork and does not extend to the full complexity of a coupled-channels analysis.

The present paper is organized as follows. In Sec.~\ref{Sec:formalism}, we briefly introduce the framework of our theoretical model, including the Reggeized $t$-channel amplitudes, effective Lagrangians, resonance propagators, form factors, amplitudes of $t$-channel $f_{2}$ exchange diagram, and formula for spin density matrix elements. In Sec.~\ref{Sec:results}, we present our theoretical outcomes, along with a detailed analysis of the contributions from background terms and nucleon resonances in the considered reaction. Finally, Sec.~\ref{Sec:summary} offers a summary and conclusions.

\section{Formalism}  \label{Sec:formalism}

\begin{figure}[tbp]
\subfigure[~$s$ channel]{
\includegraphics[width=0.45\columnwidth]{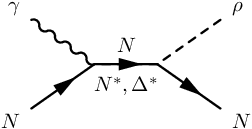}}  {\hglue 0.4cm}
\subfigure[~$t$ channel]{
\includegraphics[width=0.45\columnwidth]{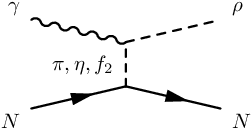}} \\[6pt]
\subfigure[~$u$ channel]{
\includegraphics[width=0.45\columnwidth]{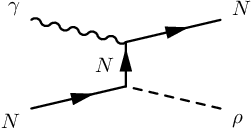}} {\hglue 0.4cm}
\subfigure[~Interaction current]{
\includegraphics[width=0.45\columnwidth]{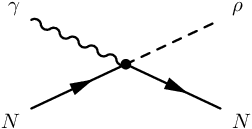}}
\caption{Generic structure of the amplitude for $\gamma N\to \rho N$. Time proceeds from left to right. }
\label{FIG:feymans}
\end{figure}

Following Refs.~\cite{Haberzettl:1997,Haberzettl:2006}, the full photoproduction amplitudes for the $\gamma N \to \rho N$ reaction can be expressed as
\begin{equation}
M^{\nu\mu} = M^{\nu\mu}_s + M^{\nu\mu}_t + M^{\nu\mu}_u + M^{\nu\mu}_{\rm int},  \label{eq:amplitude}
\end{equation}
with $\nu$ and $\mu$ being the Lorentz indices of vector meson $\rho$ and the photon $\gamma$, respectively. The $M^{\nu\mu}_s$, $M^{\nu\mu}_t$, and $M^{\nu\mu}_u$ represent the $s$-, $t$-, and $u$-channel pole diagrams, respectively, with the indices $s$, $t$, and $u$ being the Mandelstam variables of the internally exchanged particles. These three diagrams arise from the photon attaching to the external particles in the underlying $NN\rho$ interaction vertex. The term $M^{\nu\mu}_{\rm int}$ represents the interaction current, which is obtained by attaching the photon to the internal structure of the $NN\rho$ interaction vertex. The Feynman diagrams of all these four terms in Eq.~(\ref{eq:amplitude}) are depicted in Fig.~\ref{FIG:feymans}.

As shown in Fig.~\ref{FIG:feymans} that to construct the theoretical amplitudes, the contributions of the following hadron exchanges are considered: (i) $s$-channel $N$, $N^{\ast}$, and $\Delta^{\ast}$ exchanges, (ii) $t$-channel $\pi$, $\eta$, and $f_{2}$ exchanges, and (iii) $u$-channel $N$ exchange. The explicit expressions for these amplitudes can be obtained by evaluating the corresponding Feynman diagrams. Nevertheless, it is rather complicated to explicitly calculate the interaction current $M^{\nu\mu}_{\rm int}$, which obeys a highly nonlinear equation and contains diagrams with very complicated interaction dynamics. Furthermore, the introduction of phenomenological form factors makes it impossible to calculate the interaction current exactly even in principle. Following Refs.~\cite{Haberzettl:1997,Haberzettl:2006,Huang:2012,Huang:2013}, to effectively substitute the interaction current arising from the unknown parts of the underlying microscopic model, the interaction current is modeled by a generalized contact current,
\begin{equation}
M^{\nu\mu}_{\rm int} = \Gamma^\nu_{NN\rho}(q) C^\mu,  \label{eq:Mint}
\end{equation}
where $\Gamma^\nu_{NN\rho}(q)$ is the vertex function of $NN\rho$ coupling obtained from the Lagrangian of Eq.~(\ref{eq:L_NNrho}),
\begin{equation}
\Gamma^\nu_{NN\rho}(q) = - \, i \, g_{NN\rho} \left(\gamma^{\nu}-i\frac{\kappa_{NN\rho}}{2m_{N}}\sigma^{\nu\alpha}q_{\alpha}\right),
\end{equation}
with $q$ representing the four-momentum of the outgoing $\rho$ meson, and $C^\mu$ is an auxiliary current introduced to ensure the gauge invariance of the full photoproduction amplitudes of Eq.~(\ref{eq:amplitude}). Following Refs.~\cite{Haberzettl:2006,Huang:2012}, $C^\mu$ for $\gamma p \to \rho^0 p$ is denoted as
\begin{equation}   \label{eq:Cmu}
C^\mu = - Q_{N_{u}}\frac{f_{u}-\hat{F}}{u-p'^{2}}\left(2p'-k\right)^{\mu} - Q_{N_{s}}\frac{f_{s}-\hat{F}}{s-p^2}\left(2p+k\right)^{\mu},
\end{equation}
with
\begin{equation} \label{eq:Fhat-Kstp}
\hat{F} = 1 - \hat{h} \left(1 -  f_u\right) \left(1 - f_s\right),
\end{equation}
where $p$, $p'$, and $k$ are four-momenta of the incoming $N$, outgoing $N$, and incoming photon, respectively; $Q_{N_{u}(N_{s})}$ is the electric charge of the $u(s)$-channel $N$; $f_{u(s)}$ is the phenomenological form factor for $u(s)$-channel $N$ exchange; $\hat{h}$ is an arbitrary function except for that it should go to unity in the high-energy limit to prevent the ``violation of scaling behavior'' \cite{Drell:1972}. Following our previous works \cite{Zhangy:2021,Wei:2022,Wang:2022,Wei:2020,Wangx:2020,Zhangyx:2021,Wein:2021,Weinc:2019,Wang:2017,Wang:2018}, the value of $\hat{h}$ is taken as $\hat{h}=1$ for simplicity.

\subsection{Reggeized $t$-channel amplitudes} \label{subsec:Regge}

In the higher energy region, the effects of high spin meson exchanges need to be properly considered. This is achieved by substituting the $t$-channel Feynman amplitudes with Regge amplitudes, which are more appropriate for describing the dynamics of particle exchanges in this energy regime. In Regge theory, the exchanged particles are represented by their Regge trajectories, which capture the relationship between their spin and mass squared. These trajectories describe the collective contribution of a family of mesons with varying spins, offering a unified framework to account for their exchanges. The Regge propagator depends on the momentum transfer square $t$ and energy square $s$, and becomes particularly relevant at high energies where exchanges predominantly involve higher-spin mesons.

A notable feature of Regge theory is its ability to capture the strong degeneracy observed among the trajectories of meson families, where multiple mesons with similar quantum numbers align along the same trajectory. This approach provides a more accurate description of the scattering amplitude at high energies by effectively incorporating the contributions of these higher-spin particles.
	
The adoption of Reggeization enables a realistic modeling of meson exchanges in hadron production reactions, as has been demonstrated in numerous previous studies \cite{Guidal:1997hy,Vanderhaeghen:1997ts,Guidal:2003qs,He:2010ii,Mathieu:2015eia,Corthals:2005ce,DeCruz:2012bv,Vrancx:2014pwa,Chiang:2002vq,Vancraeyveld:2009qt,Kaskulov:2010kf,Yu:2011zu,Bydzovsky:2019hgn}.

The standard Reggeization of the $t$-channel $\pi$, $\eta$, and $f_{2}$ exchanges corresponds to the following replacements:
\begin{align}
\frac{1}{t-m^2_\pi}  \quad  \Longrightarrow \quad  & {\cal P}^\pi_{\rm Regge} =  \left(\frac{s}{s_0}\right)^{\alpha_\pi(t)} e^{-i\pi\alpha_\pi(t)} \nonumber \\
& \times  \frac{\pi \alpha'_\pi}{\sin[\pi\alpha_\pi(t)]} \frac{1}{\Gamma[1+\alpha_\pi(t)]}, \label{eq:Regge_prop_pi}  \\[6pt]
\frac{1}{t-m^2_\eta}  \quad \Longrightarrow \quad & {\cal P}^\eta_{\rm Regge} =  \left(\frac{s}{s_0}\right)^{\alpha_\eta(t)} e^{-i\pi\alpha_\eta(t)} \nonumber \\
& \times  \frac{\pi \alpha'_\eta}{\sin[\pi\alpha_\eta(t)]} \frac{1}{\Gamma[1+\alpha_\eta(t)]}, \label{eq:Regge_prop_eta}  \\[6pt]
\frac{1}{t-m^2_{f_{2}}}  \quad \Longrightarrow \quad & {\cal P}^{f_{2}}_{\rm Regge} = \left(\frac{s}{s_0}\right)^{\alpha_{f_{2}}(t)-2} \frac{1 + e^{-i\pi\alpha_{f_{2}}(t)}}{2} \nonumber \\
& \times  \frac{\pi \alpha'_{f_{2}}}{\sin[\pi\alpha_{f_{2}}(t)]} \frac{1}{\Gamma[\alpha_{f_{2}}(t)-1]}. \label{eq:Regge_prop_f2}
\end{align}
Here $s_0$ is a mass scale which is conventionally chosen as $s_0=1$ GeV$^2$; $\alpha'_M$ is the slope of the Regge trajectory $\alpha_M(t)$. The trajectories for $M=\pi$, $\eta$, and $f_{2}$ exchanges are respectively parametrized as \cite{Yu:2017,Mathieu:2018,Irving:1977}
\begin{align}
\alpha_{\pi}(t) &= 0.7\, {\rm GeV}^{-2} \left(t -m_{\pi}^2\right), \label{eq:trajectory_pi}  \\[3pt]
\alpha_{\eta}(t) &= 0.7\, {\rm GeV}^{-2} \left(t -m_{\eta}^2\right),  \label{eq:trajectory_eta}  \\[3pt]
\alpha_{f_{2}}(t) &= 2 + 0.9\, {\rm GeV}^{-2} \left(t -m_{f_{2}}^2\right). \label{eq:trajectory_f2}
\end{align}

\subsection{Effective Lagrangians} \label{Sec:Lagrangians}

The effective interaction Lagrangians used to construct the production amplitudes for all considered exchanges, except for the $t$-channel $f_{2}$ exchange for which the amplitude is introduced in Sec.~\ref{Sec:amplitude_f2}, are presented in this section. In order to make the following Lagrangians simple, we define the operators
\begin{equation}
\Gamma^{(+)}=\gamma_5  \qquad  \text{and} \qquad  \Gamma^{(-)}=1,
\end{equation}
and the field-strength tensors for photon and $\rho$ vector meson
\begin{align}
F^{\mu\nu} &= \partial^{\mu}A^\nu-\partial^{\nu}A^\mu,   \\[3pt]
{\rho}^{\mu\nu} &= \partial^{\mu}{\rho}^\nu-\partial^{\nu}{\rho}^\mu,
\end{align}
where $A^\mu$ and ${\rho}^{\mu}$ represent the electromagnetic field and $\rho$ meson field, respectively.

The electromagnetic interaction Lagrangians required to calculate the Feynman diagrams of the background terms are
\begin{align}
{\cal L}_{NN\gamma} &= -e \bar{N} \left[ \left( \hat{e} \gamma^\mu - \frac{ \hat{\kappa}_N} {2M_N}\sigma^{\mu \nu}\partial_\nu\right) A_\mu\right] N, \\[6pt]
{\cal L}_{\pi\rho\gamma } &= e \frac{g_{\pi\rho\gamma}}{M_{\pi}}\varepsilon^{ \alpha \mu \lambda \nu}\left(\partial_{\alpha }A_{\mu}\right) \left(\partial_{\lambda}\pi\right)\rho_{\nu},   \\[6pt]
{\cal L}_{\eta\rho\gamma } &= e \frac{g_{\eta\rho\gamma}}{M_{\eta}}\varepsilon^{ \alpha \mu \lambda \nu} \left(\partial_{\alpha }A_{\mu}\right) \left(\partial_{\lambda}\eta\right) \rho_{\nu},
\end{align}
where $e$ is the elementary charge unit, and $\hat{e}$ represents the charge operator. The anomalous magnetic moments are defined as $\hat{\kappa}_N = \kappa_p(1+\tau_3)/2 + \kappa_n(1-\tau_3)/2$, with the anomalous magnetic moments $\kappa_p=1.793$ and $\kappa_n=-1.913$. $M_N$ stands for the mass of $N$, and $\varepsilon_{\mu\nu\alpha\beta}$ is the totally antisymmetric Levi-Civita tensor with $\varepsilon^{0123}=1$. The coupling constants $g_{\eta\rho\gamma}$ and $g_{\pi\rho\gamma}$ are determined by a fit to the vector meson radiative decay width
\begin{equation}
\Gamma _{\rho\rightarrow P \gamma } = \frac{e^2}{4\pi} \frac{g_{P \rho\gamma }^{2}}{24M_{\rho}^{3}M_{p}^2} \left(M_{\rho}^{2}-M_{P}^{2}\right)^{3},
\end{equation}
where $P$ stands for the pseudoscalar meson $\pi$ or $\eta$. Using the decay width $\Gamma _{\rho^{0} \rightarrow \pi^{0} \gamma }\simeq 0.070$ MeV and $\Gamma _{\rho^{0} \rightarrow \eta \gamma }\simeq 0.045$ MeV advocated in the most recent version of Particle Data Book (PDB) \cite{PDG:2022}, one gets $|g_{\pi \rho\gamma}| \simeq 0.099$ and $|g_{\eta \rho\gamma}| \simeq 0.90$. The positive values are employed in the present work.

The effective Lagrangians for background meson-baryon interactions are
\begin{align}
{\cal L}_{NN\rho} &= -g_{NN\rho}\bar{N}\left[ \left(\gamma_{\mu }-\frac{\kappa_{NN\rho}}{2M_{N}}\sigma_{\mu \nu }\partial^{\nu }\right){\rho}^{\mu }\right]{N},  \label{eq:L_NNrho}  \\[6pt]
{\cal L}_{NN\pi} &= -g_{NN\pi}\bar{N}\gamma_5\left[ \left(i\lambda+\frac{1-\lambda}{2M_{N}}\slashed{\partial}\right)\pi \right]N,  \label{eq:L_NNpi}  \\[6pt]
{\cal L}_{NN\eta} &= -g_{NN\eta}\bar{N}\gamma_5\left[ \left(i\lambda+\frac{1-\lambda}{2M_{N}}\slashed{\partial}\right)\eta \right]N,  \label{eq:L_NNeta}
\end{align}
where the parameter $\lambda$ is introduced into ${\cal L}_{NN\pi}$ and  ${\cal L}_{NN\eta}$ to interpolate between the pseudovector $(\lambda=0)$ and the pseudoscalar $(\lambda=1)$ couplings. Following our previous work on $\gamma p \to K^{*+}\Lambda$ reaction \cite{Wang:2017}, we choose $\lambda=0$ for ${\cal L}_{NN\pi}$ and ${\cal L}_{NN\eta}$. The empirical values $g_{NN\rho}=3.25$, $\kappa_{NN\rho}=f_{NN\rho}/g_{NN\rho}=6.1$, $g_{NN\pi}=13.46$, and $g_{NN\eta}=4.76$ are quoted from Refs.~\cite{Huang:2012,Ronchen:2013}.

The resonance-nucleon-photon transition Lagrangians are
\begin{align}
{\cal L}_{RN\gamma}^{1/2\pm} =& \, e\frac{g_{RN\gamma}^{(1)}}{2M_N}\bar{R} \Gamma^{(\mp)}\sigma_{\mu\nu} \left(\partial^\nu A^\mu \right) N  + \hc, \\[6pt]
{\cal L}_{RN\gamma}^{3/2\pm} =& - i e\frac{g_{RN\gamma}^{(1)}}{2M_N}\bar{R}_\mu \gamma_\nu \Gamma^{(\pm)}F^{\mu\nu}N \nonumber \\
& + e\frac{g_{RN\gamma}^{(2)}}{\left(2M_N\right)^2}\bar{R}_\mu \Gamma^{(\pm)}F^{\mu \nu}\partial_\nu N + \hc, \\[6pt]
{\cal L}_{RN\gamma}^{5/2\pm} = & \, e\frac{g_{RN\gamma}^{(1)}}{\left(2M_N\right)^2}\bar{R}_{\mu \alpha}\gamma_\nu \Gamma^{(\mp)}\left(\partial^{\alpha} F^{\mu \nu}\right)N \nonumber \\
& \pm ie\frac{g_{RN\gamma}^{(2)}}{\left(2M_N\right)^3}\bar{R}_{\mu \alpha} \Gamma^{(\mp)}\left(\partial^\alpha F^{\mu \nu}\right)\partial_\nu N + \hc,
\end{align}
where $R$ designates the nucleon resonance, and the superscript of ${\cal L}_{RN\gamma}$ denotes the spin and parity of the resonance $R$. The coupling constants $g_{RN\gamma}^{(i)}$ $(i=1,2)$ are fit parameters.

The effective Lagrangians for hadronic vertices including nucleon resonances are
\begin{align}
{\cal L}_{RN\rho}^{1/2\pm} =& - \frac{g_{RN\rho}}{2M_N}\bar{R}\Gamma^{(\mp)} \left\{ \left[ \left(
\frac{\gamma_\mu\partial^2}{M_R\mp M_N} \pm i\partial_\mu \right)  \right.\right. \nonumber \\
&  - \! \left.\left. \frac{f_{RN\rho}}{g_{RN\rho}}\sigma_{\mu\nu}\partial^\nu \right] {\rho}^\mu
\right\} N   + \hc, \\[6pt]
{\cal L}_{RN\rho}^{3/2\pm} =& - i\frac{g_{RN\rho}^{(1)}}{2M_N}\bar{R}_\mu \gamma_\nu \Gamma^{(\pm)}{\rho}^{\mu \nu} N  \nonumber \\
& + \frac{g_{RN\rho}^{(2)}}{\left(2M_N\right)^2}\bar{R}_\mu \Gamma^{(\pm)}{\rho}^{\mu \nu}\partial_\nu N  \nonumber \\
& \mp \frac{g_{RN\rho}^{(3)}}{\left(2M_N\right)^2}\bar{R}_\mu \Gamma^{(\pm)}\left(\partial_\nu {\rho}^{\mu \nu}\right) N   + \hc, \\[6pt]
{\cal L}_{RN\rho}^{5/2\pm} = & \, \frac{g_{RN\rho}^{(1)}}{\left(2M_N\right)^2}\bar{R}_{\mu \alpha}\gamma_\nu \Gamma^{(\mp)}\left(\partial^{\alpha} {\rho}^{ \mu \nu}\right) N   \nonumber \\
& \pm  i\frac{g_{RN\rho}^{(2)}}{\left(2M_N\right)^3}\bar{R}_{\mu \alpha} \Gamma^{(\mp)}\left(\partial^\alpha {\rho}^{\mu \nu}\right)\partial_\nu N  \nonumber \\
& \mp  i\frac{g_{RN\rho}^{(3)}}{\left(2M_N\right)^3}\bar{R}_{\mu \alpha} \Gamma^{(\mp)} \left(\partial^\alpha \partial_\nu {\rho}^{\mu \nu}\right) N + \hc
\end{align}
In the present work, the coupling constant $f_{RN\rho}$ in ${\cal L}_{RN\rho}^{1/2\pm}$ is set to be zero, and the $g_{RN\rho}^{(2)}$ and $g_{RN\rho}^{(3)}$ terms in ${\cal L}_{RN\rho}^{3/2\pm}$ and ${\cal L}_{RN\rho}^{5/2\pm}$ are ignored for the sake of simplicity. Actually, these terms have been checked and found to be insensitive to the reaction amplitudes in the present investigation. The parameters $g_{RN\rho}$ and $g_{RN\rho}^{(1)}$ are fit parameters. In the fitting processes, the reaction amplitudes are relevant to the products of the electromagnetic couplings and the hadronic couplings of nucleon resonances, and thus, these products are effectively fitted in practice.

\subsection{Resonance propagators}

For spin-$1/2$ resonance propagator, we use the ansatz
\begin{equation}
S_{1/2}(p) = \frac{i}{\slashed{p} - M_R + i \Gamma/2},  \label{eq:prop-1hf}
\end{equation}
where $M_R$ and $\Gamma$ are the mass and width of resonance $R$, respectively, and $p$ is the four-momentum of the resonance $R$.

Following Refs.~\cite{Behrends:1957,Fronsdal:1958,Zhu:1999}, the prescriptions of the propagators for resonances with spin-$3/2$ and -$5/2$ are
\begin{align}
S_{3/2}(p) =& \,  \frac{i}{\slashed{p} - M_R + i \Gamma/2} \left( \tilde{g}_{\mu \nu} + \frac{1}{3} \tilde{\gamma}_\mu \tilde{\gamma}_\nu \right),  \\[6pt]
S_{5/2}(p) =& \, \frac{i}{\slashed{p} - M_R + i \Gamma/2} \,\bigg[ \, \frac{1}{2} \big(\tilde{g}_{\mu \alpha} \tilde{g}_{\nu \beta} + \tilde{g}_{\mu \beta} \tilde{g}_{\nu \alpha} \big)  \nonumber \\
& - \frac{1}{5}\tilde{g}_{\mu \nu}\tilde{g}_{\alpha \beta}  + \frac{1}{10} \big(\tilde{g}_{\mu \alpha}\tilde{\gamma}_{\nu} \tilde{\gamma}_{\beta} + \tilde{g}_{\mu \beta}\tilde{\gamma}_{\nu} \tilde{\gamma}_{\alpha}  \nonumber \\
& + \tilde{g}_{\nu \alpha}\tilde{\gamma}_{\mu} \tilde{\gamma}_{\beta} +\tilde{g}_{\nu \beta}\tilde{\gamma}_{\mu} \tilde{\gamma}_{\alpha} \big) \bigg],  \label{eq:prop-5hf}
\end{align}
where
\begin{align}
\tilde{g}_{\mu \nu} &= - \, g_{\mu \nu} + \frac{p_{\mu} p_{\nu}}{M_R^2}, \\[6pt]
\tilde{\gamma}_{\mu} &= \gamma^{\nu} \tilde{g}_{\nu \mu} = -\gamma_{\mu} + \frac{p_{\mu}\slashed{p}}{M_R^2}.
\end{align}

As shown in Eqs.~\eqref{eq:prop-1hf}-\eqref{eq:prop-5hf}, in the present work, we employ the Rarita-Schwinger prescription of resonance propagators. This choice provides an economic and convenient approximation while maintaining consistency with prior studies. However, we acknowledge the presence of unphysical components in these propagators and recognize the importance of rigorous treatments, such as those discussed in Ref.~\cite{Mart:2019jtb}. These advanced treatments will be considered in future studies when more data for this reaction become available.

\subsection{Form factors}\label{Sec:formfactor}

Each hadronic vertex obtained from the Lagrangians given in Sec.~\ref{Sec:Lagrangians} is accompanied by a phenomenological form factor to regularize the production amplitude and parametrize the internal structures of the interacting hadrons. In particular, the form factor $f_B$ in the $s$-channel is introduced to ensure compliance with the Froissart bound, which limits the high-energy growth of the scattering amplitude. This regularization prevents unphysical contributions at high energies and stabilizes the theoretical predictions, ensuring the physical consistency of the model. Following Refs.~\cite{Wang:2017,Wang:2018}, for intermediate baryon exchange, we take the form factor as
\begin{eqnarray}
f_B(p^2) = \left(\frac{\Lambda_B^4}{\Lambda_B^4+\left(p^2-M_B^2\right)^2}\right)^2,  \label{eq:ff_B}
\end{eqnarray}
where $p$ denotes the four-momentum of the intermediate baryon, and $M_B$ is the mass for exchanged baryon $B$. The cutoff $\Lambda_B$ is treated as a fitting parameter for each exchanged baryon.
For $t$-channel meson exchange, we take the form factor as
\begin{eqnarray}
f_M(q^2) = \left(\frac{\Lambda_M^2-M_M^2}{\Lambda_M^2-q^2}\right)^2, \label{eq:ff_M}
\end{eqnarray}
where $q$ represents the four-momentum of the intermediate meson, and $M_M$ and $\Lambda_M$ designate the mass and cutoff mass of exchanged meson $M$, respectively.

\subsection{Amplitude of $t$-channel $f_2$ exchange} \label{Sec:amplitude_f2}

Following Refs.~\cite{Gold68,PSMM73,Yu:2019,Achuthan:1970}, the effective Lagrangian for the $NNf_{2}$ coupling is denoted as
\begin{align}
{\cal L}_{NNf_{2}} = & \, 2i\frac{G_{NNf_{2}}}{M_N} \bar{N} \left(\gamma_\mu\partial_\nu + \gamma_\nu \partial_\mu \right) N f^{\mu\nu} \nonumber \\
& + 4 \frac{F_{NNf_{2}}}{M_N^2} \left(\partial_\mu \bar{N} \right) \left(\partial_\nu N \right) f^{\mu\nu},
\end{align}
where $f^{\mu\nu}$ is the $f_2$ meson field. The $NNf_{2}$ vertex function can then be written as
\begin{align}
\Gamma_{NNf_{2}}^{\alpha\beta} =& \; i \bigg\{ \frac{G_{NNf_{2}}}{M_N} \left[ \left(p+p'\right)^{\alpha} \gamma^\beta + \gamma^\alpha \left(p+p'\right)^{\beta} \right]    \nonumber \\[3pt]
&  + \frac{F_{NNf_{2}}}{M_N^2} \left(p+p'\right)^{\alpha} \left(p+p'\right)^{\beta} \bigg\},
\end{align}
with $p$ and $p'$ being the four-momentum of incoming and outgoing nucleons, respectively. Following Refs.~\cite{Yu:2019,Achuthan:1970}, the couplings $G_{NNf_{2}}$ and $F_{NNf_{2}}$ are set to $G_{NNf_{2}}=6.45$ and $F_{NNf_{2}}=0$.

The gauge-invariant vertex function for $f_{2}\rho\gamma$ coupling can be written as \cite{Renn:71}
\begin{align}
\Gamma^{\rho\sigma;\mu\nu}_{f_{2}\rho\gamma}(k,q) =& \; i \frac{g_{f_{2}\rho\gamma}^{}}{M_{f_{2}}} \, \big[ \, g^{\mu\nu} \left(k+q\right)^\rho \left(k+q\right)^\sigma  \nonumber \\[3pt]
& - \left( g^{\nu\sigma} q^\mu + g^{\mu\sigma} k^\nu \right) \left(k+q\right)^\rho \nonumber \\[3pt]
& - \left( g^{\mu\rho} k^\nu + g^{\nu\rho} q^\mu \right) \left(k+q\right)^\sigma \nonumber \\[3pt]
&  + 2 k \cdot q \left( g^{\mu\rho} g^{\nu\sigma} + g^{\mu\sigma} g^{\nu\rho} \right) \big],
\end{align}
where $k$ and $q$ are four-momentum of incoming photon and outgoing $\rho$ meson, respectively, and  \cite{Oh:2004,Yu:2019}
\begin{equation}
g_{f_{2}\rho\gamma} = \frac{e}{f_{\rho}}G_{f_{2}\rho\rho} = \frac{e}{5.2} \times 5.76.
\end{equation}
Thus, the amplitude for the $t$-channel $f_{2}$ exchange can be written as
\begin{align}
{M}^{\nu\mu}_{f_2} =&\;  \bar{u}(p') \Gamma^{\alpha\beta}_{NNf_{2}}(p,p') u(p) \, i P_{\alpha\beta;\rho\sigma} {\cal P}^{f_{2}}_{\rm Regge} \nonumber \\[3pt]
& \times \Gamma^{\rho\sigma;\mu\nu}_{f_{2}\rho\gamma}(k,q)f_{f_{2}}(t),
\end{align}
where $f_{f_{2}}(t)$ is the form factor attached to the amplitude of $f_{2}$ exchange, as introduced in Sec.~\ref{Sec:formfactor}.
In addition, the $P_{\alpha\beta;\rho\sigma}$ for the $f_{2}$ tensor exchange is taken as
\begin{equation}
P_{\alpha\beta;\rho\sigma} = \frac{1}{2} \left(\bar{g}_{\alpha\rho} \bar{g}_{\beta\sigma} + \bar{g}_{\alpha\sigma} \bar{g}_{\beta\rho}\right) - \frac13 \bar{g}_{\alpha\beta} \bar{g}_{\rho\sigma},
\end{equation}
with
\begin{equation}
\bar{g}_{\mu\nu} = - g_{\mu\nu} + \frac{(p-p')_\mu (p-p')_\nu}{M_{f_2}^2}.
\end{equation}

\subsection{Observables}

To define the relevant observables, we establish a set of three mutually orthogonal unit vectors $\{\hat{x}, \hat{y}, \hat{z}\}$ based on the available momenta, specifically the incident photon momentum $\mathbf{k}$ and the outgoing $\rho$-meson momentum $\mathbf{q}$:
\begin{equation}
\hat{z} \equiv \frac{\mathbf{k}}{|\mathbf{k}|}, \qquad \hat{y} \equiv \frac{\mathbf{k} \times \mathbf{q}}{|\mathbf{k} \times \mathbf{q}|}, \qquad \hat{x} \equiv \hat{y} \times \hat{z}. \label{eq:refframe}
\end{equation}
Here, boldface notation denotes three-momentum vectors.

In the center-of-mass (c.m.) frame of the $\gamma p \to \rho^0 p$ reaction, the invariant reaction amplitude $M^{\nu\mu}$, introduced in the previous subsection [see Eq.~(\ref{eq:amplitude})], can be expressed in the c.m. helicity basis \cite{Jacob:1964}:
\begin{equation}
T_{\lambda_\rho\lambda_f, \lambda_\gamma\lambda_i} \equiv \Braket{\mathbf{q},\lambda_\rho; \mathbf{p}_f,\lambda_f | M | \mathbf{k},\lambda_\gamma; \mathbf{p}_i,\lambda_i}, \label{eq:HelMtrx}
\end{equation}
where $\lambda_\gamma$, $\lambda_\rho$, $\lambda_i$, and $\lambda_f$ represent the helicities of the photon, $\rho$-meson, initial nucleon, and final nucleon, respectively. The normalization of the helicity amplitude ensures its relationship with the differential cross section:
\begin{equation}
\frac{d\sigma}{d\Omega} = \frac{1}{64\pi^2 s} \frac{|\mathbf{q}|}{|\mathbf{k}|} \times \frac{1}{4} \sum_{\lambda_\gamma\lambda_\rho\lambda_i\lambda_f} \left| T_{\lambda_\rho\lambda_f, \lambda_\gamma\lambda_i} \right|^2. \label{eq:3-1}
\end{equation}

Following the formalism of Ref.~\cite{Fasano:1992es}, the cross section $d\sigma/d\Omega$ is denoted by $\sigma(B, T; R, V)$, where the arguments $(B, T; R, V)$ refer to the polarization states of the photon beam ($B$), target proton ($T$), recoil proton ($R$), and $\rho$-meson ($V$). For the unpolarized case, the differential cross section is
\begin{equation}
\frac{d\sigma}{d\Omega} = \sigma(U, U; U, U), \label{eq:unpol-xsc}
\end{equation}
where $U$ represents an unpolarized spin state.

The photon beam asymmetry ($\Sigma$) and target nucleon asymmetry ($T$) are given by
\begin{align}
\frac{d\sigma}{d\Omega} \Sigma &= \sigma(\bot, U; U, U) - \sigma(\|, U; U, U), \\[3pt]
\frac{d\sigma}{d\Omega} T &= \sigma(U, \bot; U, U) - \sigma(U, \|; U, U), \label{eq:spin-P}
\end{align}
where $\bot$ and $\|$ indicate polarization states perpendicular and parallel to the reaction plane, respectively.

Spin density matrix elements (SDMEs) for the $\rho$-meson are defined following Refs.~\cite{Weinc:2019,Schilling:1969um} as
\begin{align}
\rho^0_{\lambda_V \lambda'_V} &= \frac{\sum_{\lambda_f \lambda_\gamma \lambda_i} T_{\lambda_V\lambda_f, \lambda_\gamma\lambda_i} {T^*_{\lambda'_V\lambda_f, \lambda_\gamma\lambda_i}}}{\sum_{\lambda_V \lambda_f \lambda_\gamma \lambda_i} \left| T_{\lambda_V\lambda_f, \lambda_\gamma\lambda_i} \right|^2}, \nonumber \\[6pt]
\rho^1_{\lambda_V \lambda'_V} &= \frac{\sum_{\lambda_f \lambda_\gamma \lambda_i} T_{\lambda_V\lambda_f, -\lambda_\gamma\lambda_i} {T^*_{\lambda'_V\lambda_f, \lambda_\gamma\lambda_i}}}{\sum_{\lambda_V \lambda_f \lambda_\gamma \lambda_i} \left| T_{\lambda_V\lambda_f, \lambda_\gamma\lambda_i} \right|^2}, \nonumber \\[6pt]
\rho^2_{\lambda_V \lambda'_V} &= \frac{i \sum_{\lambda_f \lambda_\gamma \lambda_i} \lambda_\gamma T_{\lambda_V\lambda_f, -\lambda_\gamma\lambda_i} {T^*_{\lambda'_V\lambda_f, \lambda_\gamma\lambda_i}}}{\sum_{\lambda_V \lambda_f \lambda_\gamma \lambda_i} \left| T_{\lambda_V\lambda_f, \lambda_\gamma\lambda_i} \right|^2}, \label{eq:SDMEs}
\end{align}
where $\lambda_V$ and $\lambda'_V$ are the helicities of the outgoing $\rho$-meson.

\section{Results and discussion}   \label{Sec:results}

As mentioned in Sec.~\ref{Sec:intro}, the new data on spin density matrix elements measured at a photon laboratory energy of $E_\gamma=8.5$ GeV by the GlueX Collaboration in 2023 \cite{newsdme:2020} have never been analyzed in literature. It is expected that the reported parity spin asymmetry $P_\sigma$, which is a combination of the spin density matrix elements of $\rho^1_{1-1}$ and $\rho^1_{00}$, imposes additional constraints on the relative contributions from natural parity and unnatural parity exchanges. Notably, the present $P_{\sigma}$ approaches unity at very small momentum-transfer ($-t$) regions, suggesting that at the measured energy ($E_{\gamma}=8.5$ GeV), natural parity exchanges, rather than unnatural parity exchanges, dominate at very forward angles in this reaction. In the present work, we, for the first time, perform a combined analysis of both the newly released data on spin density matrix elements \cite{newsdme:2020} and the previously published data on differential cross sections \cite{Battaglieri:2001,Wu:2005,Ballam:1972} for $\gamma p \to \rho^0 p$ using an effective Lagrangian method at the tree-level approximation. Our primary objective is to unravel the underlying reaction mechanisms and investigate potential contributions from relevant nucleon resonances.

We consider the $t$-channel $\pi$, $\eta$, and $f_2$ exchanges, $s$-channel $N$ contribution, $u$-channel $N$ exchange, and the generalized contact term as background contributions. The $t$-channel exchanges are modeled based on Regge theory, as the spin density matrix elements were measured at $E_\gamma=8.5$ GeV, a sufficiently high value necessitating the consideration of $t$-channel high-spin meson exchanges. The incorporation of $s$-channel nucleon resonances is kept to a minimum, introduced only when the background terms alone prove insufficient to adequately describe the data.

\begin{table}[tbp]
\caption{\label{Table:para} Fitted values of adjustable parameters. The asterisks below resonance names denote the overall status of these resonances evaluated by the PDB \cite{PDG:2022}. The numbers in brackets below $M_R$ and $\Gamma_R$ represent the range of the corresponding quantities given by the PDB. The mass, width, and cutoff mass are in MeV.}
\begin{tabular*}{\columnwidth}{@{\extracolsep\fill}lrrrr}
\hline\hline
                          &  model I      &  model II       &   model III      &   model IV    \\
\hline
$\Lambda_{N_s}$   &$1549\pm 5$    & $1503\pm8$        &   $1530\pm5 $    &   $1421\pm9 $   \\
$\Lambda_{N_u}$   &$724\pm 10$    & $728\pm10$        &   $711\pm14 $    &   $719\pm9 $    \\
$\Lambda_{f_2}$   &$855\pm 1$     & $857\pm1$         &   $863\pm1 $     &   $870\pm1 $    \\
\hline
                        &               &  $N(2100){1/2}^+$ &  $N(2060){5/2}^-$&  $\Delta(2000){5/2}^+$      \\
                        &               &  $\ast$$\ast$$\ast$ & $\ast$$\ast$$\ast$ & $\ast$$\ast$  \\
$M_R$             &               &  $2150\pm3$       &  $2200\pm1$      & $2100\pm1$        \\
                        &               &  [$2050\sim 2150$]     & [$2030\sim 2200$]     &         \\
$\Gamma_R$      &               &  ${320\pm1}$      &  $450\pm3$       &$450\pm1$          \\
                        &               &  [$200\sim320$]     &  [$300\sim 450$] &                 \\
$\Lambda_R$       &               & $1800\pm 27$      &  $1133\pm 15$    & $1397\pm 8$       \\
$g_{RN\gamma}^{(1)}g_{RN\rho}^{(1)}$&   & $-1.71\pm 0.03$    &  $17.1\pm 3$    & $0.0\pm 6$        \\
$g_{RN\gamma}^{(2)}g_{RN\rho}^{(1)}$&   & $-0.20\pm 1.4$    &  $-73.6\pm 8$    & $63.5\pm 6$       \\
\hline\hline
\end{tabular*}
\end{table}

\begin{figure*}[tbp]
\includegraphics[width=0.8\textwidth]{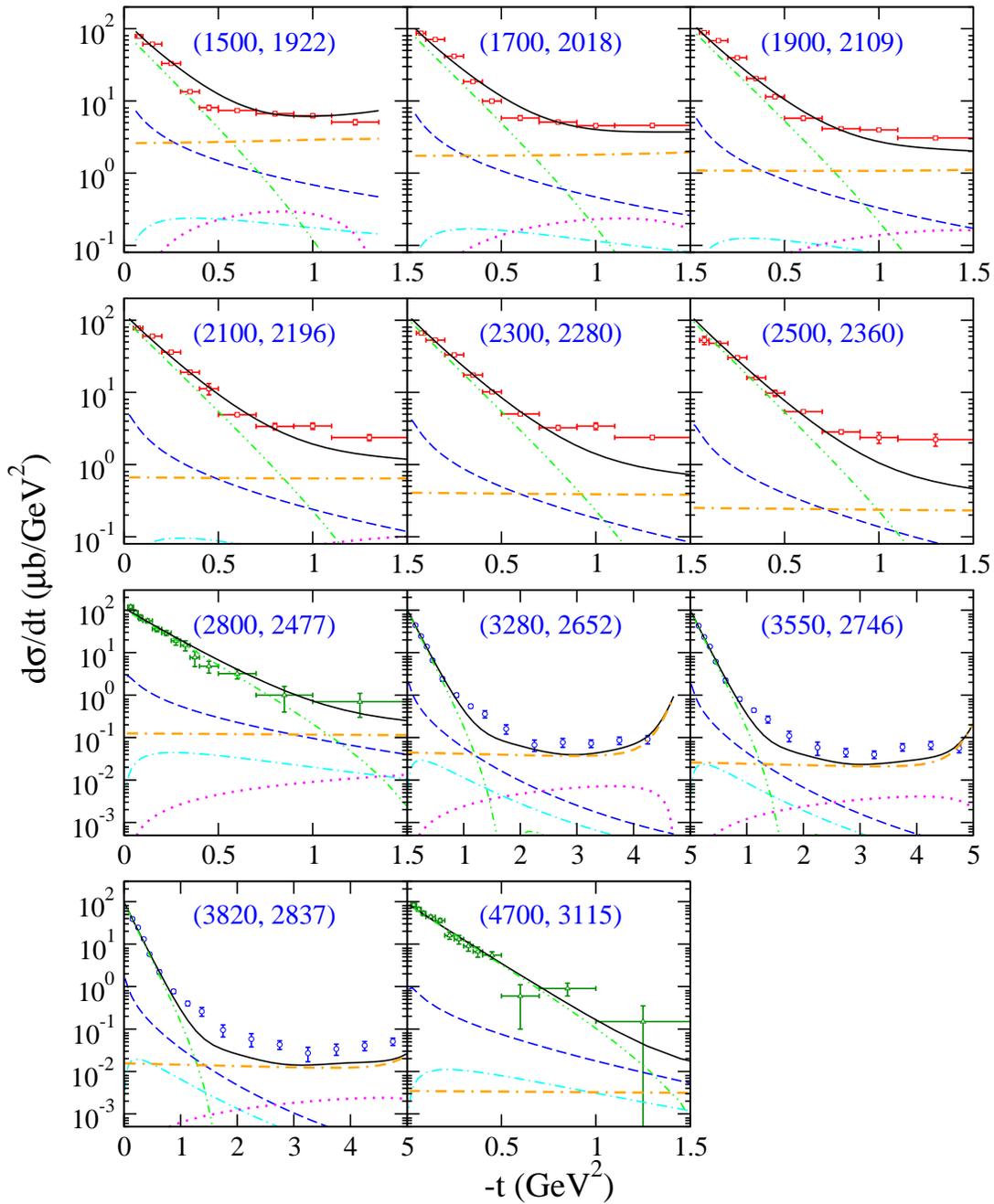}
\caption{Differential cross sections for $\gamma p \to \rho^0 p$ as a function of $-t$ obtained in model I. The solid lines represent the full results. The blue long-dashed, cyan dash-dotted, green dash-double-dotted, and orange dot-double-dashed lines represent the individual contributions from the $\pi$, $\eta$, $f_{2}$, and $N$ exchanges, respectively. The magenta dotted lines represent the individual contributions from the interaction current. The scattered red empty squares, blue empty circles, and green empty triangles denote the data from  Refs.~\cite{Battaglieri:2001}, \cite{Wu:2005}, and \cite{Ballam:1972}, respectively. The numbers in parentheses denote the centroid value of the photon laboratory incident energy (left number) and the corresponding total center-of-mass energy of the system (right number), in MeV. }
\label{fig:dif_crs}
\end{figure*}

\begin{figure*}[tbp]
\includegraphics[width=0.8\textwidth]{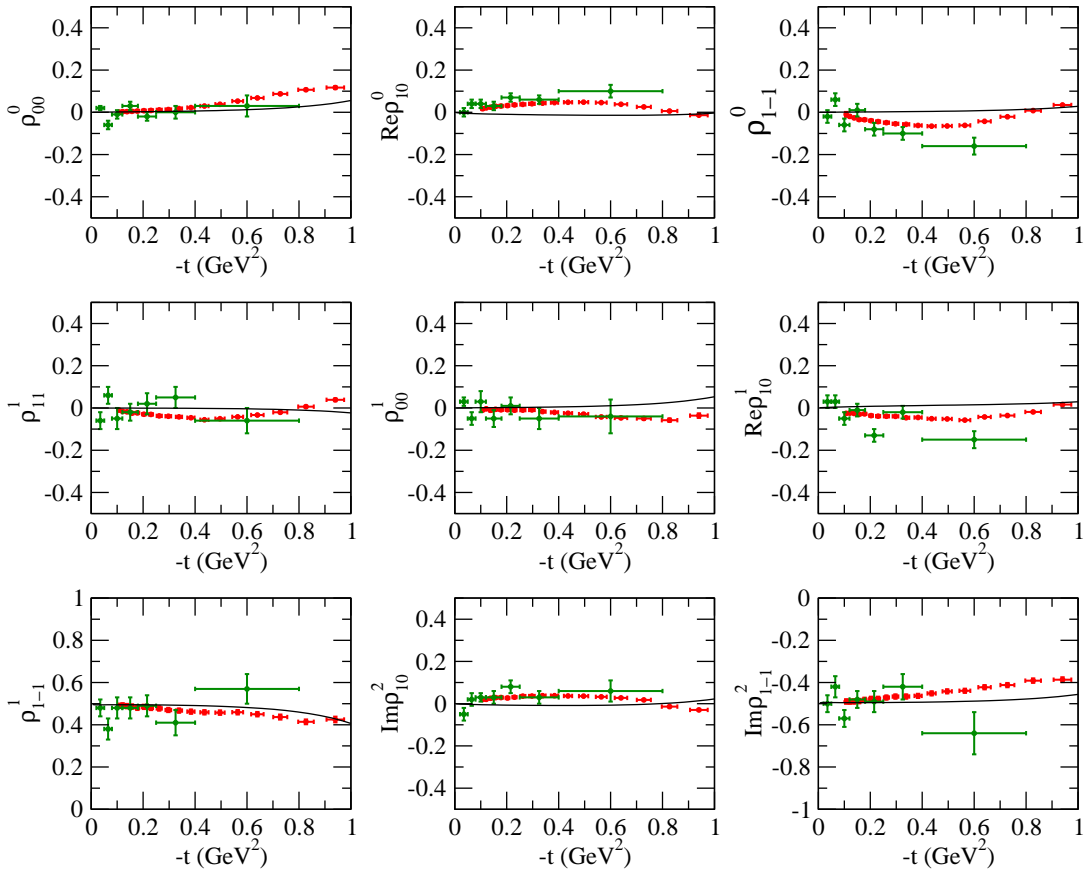}
\caption{Spin density matrix elements for $\gamma p \to \rho^0 p$ as a function of $-t$ at $E_{\gamma}=8.5$ GeV obtained in model I. The scattered red full circles and green full circles represent the data from Refs.~\cite{newsdme:2020} and \cite{oldsdme:1973}, respectively. Note that the data from Ref.~\cite{oldsdme:1973} are not included in our fit.}
\label{fig:sdme}
\end{figure*}

\begin{figure}[tbp]
\includegraphics[width=0.8\columnwidth]{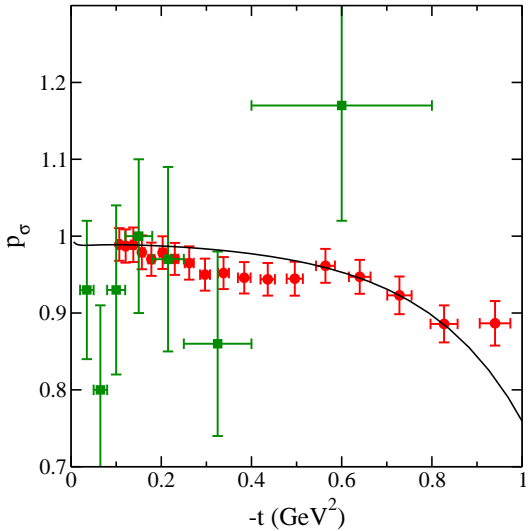}
\caption{Parity spin asymmetry $(P_{\sigma} = 2\rho^{1}_{1-1} - \rho^{1}_{00})$ for $\gamma p \to \rho^0 p$ as a function of $-t$ at $E_{\gamma}=8.5$ GeV obtained in model I. The scattered red full circles and green full squares represent the data from Refs.~\cite{newsdme:2020} and \cite{oldsdme:1973}, respectively. Note that the data from Ref.~\cite{oldsdme:1973} are not included in our fit. }
\label{fig:psig}
\end{figure}

\begin{figure*}[tbp]
\includegraphics[width=0.8\textwidth]{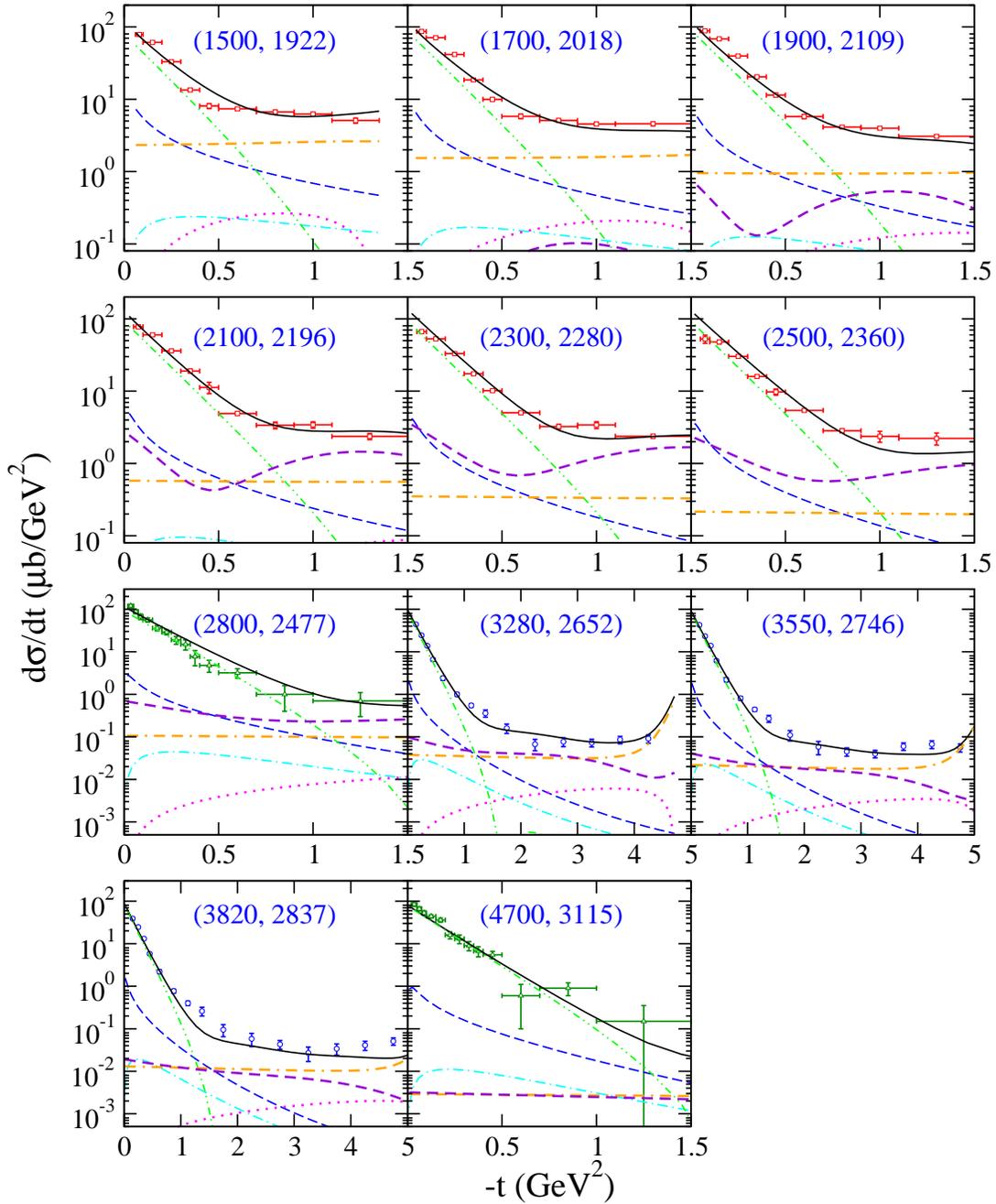}
\caption{Differential cross sections for $\gamma p \to \rho^0 p$ as a function of $-t$ for model III. The notations are the same as in Fig.~\ref{fig:dif_crs}. In addition, the violet short dashed lines represent the individual contributions from the $N(2060)5/2^-$ exchange. }
\label{fig:diff_crs_5m}
\end{figure*}

\begin{figure*}[tbp]
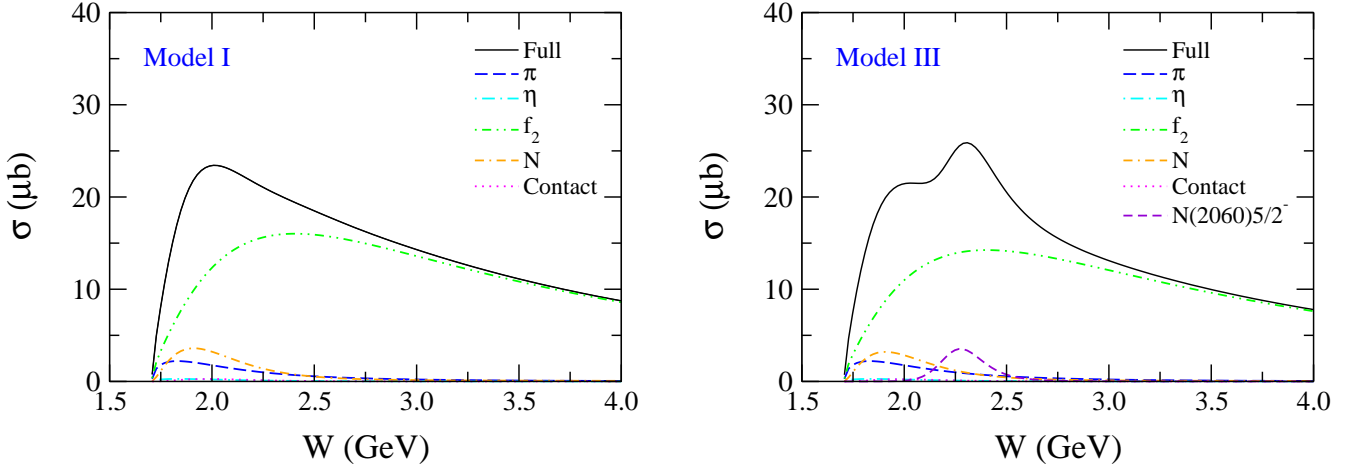

\vglue 0.1cm
\includegraphics[width=0.47\textwidth]{totl}
\hglue 0.7cm
\includegraphics[width=0.47\textwidth]{totl_5m}
\caption{Total cross sections with dominant individual contributions for $\gamma p \to \rho^0 p$. The left graph is for model I and the right one corresponds to model III. The solid lines represent the full results. The blue long dashed, cyan dash-dotted, green dash-double-dotted, orange dot-double-dashed, and magenta dotted lines represent the individual contributions from the $\pi$, $\eta$, $f_{2}$, $N$, and contact exchanges, respectively. The violet short dashed line in the right graph represents the individual contributions from the $N(2060)5/2^-$ resonance.  }
\label{fig:total_cro_sec}
\end{figure*}

\begin{figure}[tb]
\includegraphics[width=\columnwidth]{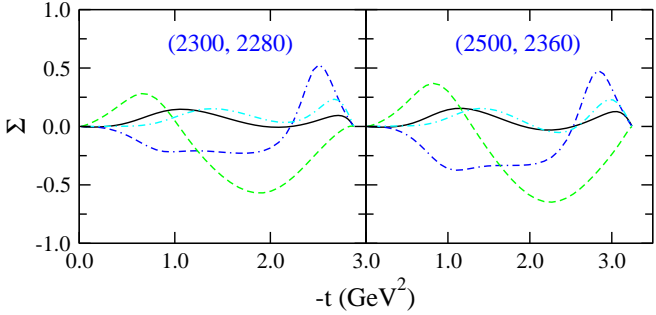}
\caption{Photon beam asymmetries for $\gamma p \to \rho^0 p$ as functions of $-t$. The cyan dot-dashed, black solid, blue dot-double-dashed, and green dashed lines represent the results from models I-IV, respectively.}
\label{fig:beam_asy}
\end{figure}

\begin{figure}[tb]
\includegraphics[width=\columnwidth]{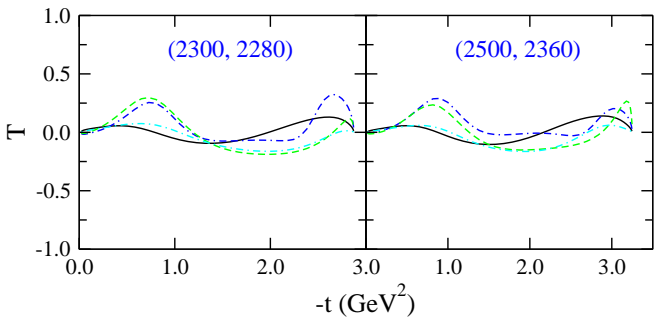}
\caption{Same as in Fig.~\ref{fig:beam_asy} for target nucleon asymmetries.}
\label{fig:target_asy}
\end{figure}

We have also checked the effects of $t$-channel $\eta'$ exchange and the contributions from Pomeron exchange in the present work. It is found that the $\eta'$ exchange contributes negligibly to the fitting results. Considering that including this term requires an additional fitting parameter for the coupling constant $g_{NN\eta'}$, we simply leave the $\eta'$ exchange out of account in the later fits. The introduction of Pomeron exchange is found to significantly worsen the describing quality of the data on spin density matrix elements, and thus this term is not included in the present work, either.

Firstly, we focus on reproducing the data without considering any nucleon resonances. After rigorous exploration, we get a qualitative description of the analyzed data by considering only the background terms. This model is named as ``model I". There are three adjustable parameters in this model, namely the cutoff parameters for $s(u)$-channel $N$ exchange $\Lambda_{{N_s}{(N_u)}}$ and for $t$-channel $f_2$ meson exchange $\Lambda_{f_2}$. The fitted values of these three parameters are listed in the second column of Table~\ref{Table:para}, where the uncertainties stem from the error bars associated with the fitted experimental data points. The corresponding results of differential cross sections, spin density matrix elements, and the parity spin asymmetry for $\gamma p\to \rho^0 p$ obtained in this model are presented in Figs.~\ref{fig:dif_crs}-\ref{fig:psig}, respectively.

Figure~\ref{fig:dif_crs} shows the differential cross section results from model I. In this figure, the black solid lines represent the full results; the blue long-dashed, cyan dash-dotted, green dash-double-dotted, and orange dot-double-dashed lines represent the individual contributions from the $\pi$, $\eta$, $f_{2}$, and $N$ exchanges, respectively; the magenta dotted lines represent the individual contributions from the interaction current. Contributions from other terms are too small to be clearly seen with the scale used, and thus, they are not plotted. The numbers in parentheses denote the centroid value of the photon laboratory incident energy (left number) and the corresponding total center-of-mass energy of the system (right number), in MeV.

One sees from Fig.~\ref{fig:dif_crs} that our overall description of the differential cross-section data is fairly good, except that the data in the energy range $E_{\rm cm} \approx 2.1-2.4$ GeV at large $-t$ values are underestimated. The individual contributions show that the $t$-channel $f_2$ exchange, followed by the $\pi$ exchange, is responsible for the sharp lifts of the angular distributions in the very small $-t$ region, while the $N$ exchange plays a significant role in the flat angular distribution in the middle $-t$ region. The $\eta$ exchange and interaction current contribute little to this reaction.

As shown in Fig.~\ref{fig:dif_crs}, the differential cross sections at $-t\approx0.5$ GeV$^2$ for the lowest two energy bins are slightly over estimated. This might indicate that steeper slope parameters are needed for the exchanged Regge trajectories. We have tried to refit the slope parameters of the trajectories to better describe the differential cross-section data, especially those at $-t\approx0.5$ GeV$^2$ in the lowest two energy bins. The results show that the fitted values of the slope parameters change a little bit, but nevertheless the differences of the calculated differential cross sections compared with those shown in Fig.~\ref{fig:dif_crs} are too tiny to be obviously noticed. In particular, the description of the data at $-t\approx 0.5$ GeV$^2$ in the lowest two energy bins cannot be improved at all. This can be understood as the Regge trajectories are  determined by the cross-section data at very forward angles in the high-energy region. In view of this, one sees that the slope parameters for considered trajectories adopted in the present work [see Eqs.~\eqref{eq:trajectory_pi}-\eqref{eq:trajectory_f2}] are reasonable and compatible with the available data.

Additionally, we have studied the influence of the relative phases of the amplitudes of $f_2$- and $\pi$-trajectory exchanges and also of $f_2$-trajectory and $N$ exchanges on differential cross sections. The results show that by choosing opposite phases for $f_2$- and $\pi$-trajectory exchanges, the changes of the cross sections are negligible at small $-t$ and a little bit worse at larger $-t$ values. When choose opposite phases for the amplitudes of $f_2$-trajectory and $N$ exchanges, the slope of the differential cross sections at small $-t$ does agree better with the cross-section data, but nevertheless, at larger $-t$ values, the description of the data get worse. Since in the present work we are seeking for an overall description of all the available data instead of the data in small $-t$ region, we prefer not choosing additional opposite phases for the amplitudes of $\pi$-trajectory and $N$ exchanges compared with that of dominant $f_2$-trajectory exchange.

Figure~\ref{fig:sdme} shows the results of spin density matrix elements from model I. One sees that the theoretical results agree only roughly with the data, while the detailed structures exhibited in the $-t$ dependence of the data on spin density matrix elements are not well described. The deviations suggest that additional contributions or refinements in the modeling of spin-dependent terms may be required to fully capture the data's complexity.

Figure~\ref{fig:psig} shows the results of parity spin asymmetry $P_{\sigma}=2\rho^{1}_{1-1}-\rho^{1}_{00}$ from model I. In the high-energy and forward-angle region where $t$-channel meson exchanges dominate the interactions, $P_{\sigma}$ tends to $1$ for natural parity exchange and $-1$ for unnatural parity exchange. From Fig.~\ref{fig:psig} one sees that our theoretical results of parity spin asymmetry $P_{\sigma}$ agree quite well with the corresponding data from the GlueX Collaboration \cite{newsdme:2020}, and the $P_{\sigma}$ approaches to unity at forward angles, indicating that the natural parity exchange dominates the contributions to this reaction in the high energy region. This is consistent with that we have seen from Fig.~\ref{fig:dif_crs} that the $f_2$ meson exchange plays the dominant role to the differential cross sections of the $\gamma p\to \rho^0 p$ reaction at forward angles.

Although the data on differential cross sections, spin density matrix elements, and parity spin asymmetry for $\gamma p \to \rho^0 p$ can be qualitatively described by considering only the background terms, the differential cross-section data are underestimated in the energy region $E_{\rm cm} \approx 2.1-2.4$ GeV at large $-t$ values, as exhibited in Fig.~\ref{fig:dif_crs}. This deviation suggests a potential deficiency in accounting for contributions from nucleon resonances.

Then we further consider the contributions from nucleon resonances to the $\gamma p\to \rho^0 p$ reaction. In the energy region $E_{\rm cm} \approx 2.0-2.4$ GeV, there are three resonances with spins ranging from $1/2$ to $5/2$ that decay into the $\rho N$ channel in the most recent version of the PDB \cite{PDG:2022}, namely the $N(2100)1/2^{+}$, $N(2060)5/2^{-}$, and $\Delta(2000)5/2^{+}$ resonances. We find that the theoretical description of the $\gamma p\to \rho^0 p$ differential cross-section data in the energy region $E_{\rm cm} \approx 2.1-2.4$ GeV at large $-t$ values can be substantially improved by including any one of these three nucleon resonances. The models with the inclusion of $N(2100)1/2^{+}$, $N(2060)5/2^{-}$, and $\Delta(2000)5/2^{+}$ are named as models II-IV, respectively.  The fitted values of the adjustable parameters in these three models are presented in the 3rd-5th columns of Table~\ref{Table:para}. One sees that the fitted values of the cut-off mass of $t$-channel meson exchanges in these models are rather stable. This is because in each of these models, the $t$-channel interaction is dominated by $f_2$ exchange which is well determined by the sharp rises of differential cross sections and the unity value of $P_\sigma$ at forward angles.

The differential cross sections in the energy region $E_{\rm cm} \approx 2.1-2.4$ GeV at large $-t$ values can be much well described in models II-IV than in model I where only background terms are considered. As an illustration, we show in Fig.~\ref{fig:diff_crs_5m} the results for differential cross sections together with dominant individual contributions from various interaction terms obtained from model III where the $N(2060)5/2^-$ resonance is included. The results from models II and IV are very similar to those from model III and are not shown in additional figures. One sees that overall the data on differential cross sections in the whole energy and angle region considered are quite well reproduced. In particular, the resonance exchange  provides significant contributions at large $-t$ values in the energy region $E_{\rm cm} \approx 2.1-2.4$ GeV, serving as necessary supplements to the dominant background contributions. The effects from other individual contributions change little compared with model I.

The results of spin density matrix elements from models II-IV are almost the same as those in model I [c.f. Figs.~\ref{fig:sdme}-\ref{fig:psig}]. This can be understood if we notice that the resonance masses are in the energy region $E_{\rm cm} \approx 2.0- 2.1$ GeV that corresponds to $E_\gamma \approx 1.7- 1.9$ GeV, very far away from the energy $E_\gamma = 8.5$ GeV, where the data on spin density matrix elements and parity spin asymmetry are measured, to result in significant effects.

In Fig.~\ref{fig:total_cro_sec}, we show the results of total cross sections for $\gamma p \to \rho^0 p$ from models I (left panel) and III (right panel). It can be observed that, in both models I and III, the $t$-channel $f_{2}$ exchange contributes dominantly in the considered energy region for this reaction. The $\pi$ exchange, along with the $N$ exchange, also contributes significantly near the $\rho N$ threshold energy region. In model III, the nucleon resonance $N(2060)5/2^{-}$ contributes considerably in the energy range $E_{\rm cm}\approx 2.0-2.5$ GeV, where the description quality of differential cross sections in model I needs improvement. The $\eta$ exchange and contact term contribute little in these two models. The results of total cross sections from models II and IV are similar to those from model III.

In Figs.~\ref{fig:beam_asy}--\ref{fig:target_asy}, we show the predictions of the photon beam asymmetry ($\Sigma$) and target nucleon asymmetry ($T$) from these four models. There, the cyan dot-dashed, black solid, blue dot-double-dashed, and green dashed lines represent the results from models I-IV, respectively. One sees that although the differential cross sections and the spin density matrix elements obtained from these models are quite similar, the $\Sigma$ and $T$ predicted from them are quite different. These spin observables can be used to distinguish the theoretical models and clarify the reaction mechanisms of the $\gamma p \to \rho^0 p$ reaction. We hope that the corresponding data can be measured in future experiments.

\section{Summary and conclusion}  \label{Sec:summary}

For the first time, we perform a combined analysis of the most recent data on spin density matrix elements published by the GlueX Collaboration at a photon energy $E_{\gamma}= 8.5$ GeV \cite{newsdme:2020} and previously published data on differential cross sections \cite{Battaglieri:2001,Wu:2005,Ballam:1972} for the $\gamma p \to \rho^0 p$ reaction within an effective Lagrangian approach. The model incorporates $t$-channel $\pi$, $\eta$, and $f_2$ exchanges, $s$- and $u$-channel $N$ exchanges, a generalized contact term, and possible $s$-channel $N^\ast$ exchanges in constructing the reaction amplitudes.

Our analysis reveals that the background contributions alone, dominated by the $t$-channel exchanges and particularly the $f_2$ exchange modeled using the Regge formalism, are able to qualitatively describe most of the experimental data, especially the data at forward angles. The $f_2$-trajectory exchange shapes the angular distributions and yields the unity value of the parity spin asymmetry $P_\sigma$ at forward angles. Additionally, the $\pi$ and $N$ exchanges contribute significantly near the $\rho N$ threshold energy region, while the contributions from the $\eta$ exchange and contact term are found to be negligible.

Nevertheless, the differential cross-section data in the energy range $E_{\rm cm} \approx 2.1-2.4$ GeV at large $-t$ values are underestimated when only background terms are considered, indicating the necessity of a potential contribution from nucleon resonances. By including one of the $N(2100)1/2^+$, $N(2060)5/2^-$, or $\Delta(2000)5/2^+$ resonances, a satisfactory description of all the considered data can be achieved. These nucleon resonances play a particularly important role in improving the description of the differential cross-section data at near-perpendicular scattering angles in the energy region $E_{\rm cm}\approx 2.0-2.5$ GeV.

Predictions for the photon beam asymmetry $\Sigma$ and target nucleon asymmetry $T$ are made, which provide a means to discriminate between theoretical models and to clarify the reaction mechanisms, particularly the contributions of nucleon resonances. Future experimental data on these observables are anticipated to offer valuable insights into the underlying dynamics of the $\gamma p \to \rho^0 p$ reaction, further refining our understanding of the interplay between Regge theory and nucleon resonances.

\begin{acknowledgments}
This work is partially supported by the National Natural Science Foundation of China under Grants No.~12175240, No.~12147153, No.~12305097, and No.~12305137, the Fundamental Research Funds for the Central Universities (Grant No.~23CX06037A), the Shandong Provincial Natural Science Foundation, China (Grant No.~ZR2024QA096), and the Taishan Scholar Young Talent Program (tsqn202408091).
\end{acknowledgments}

\end{document}